\documentclass{Interspeech}

\interspeechcameraready

\title{Use Cases for Voice Anonymization}

\author{Sarina}{Meyer}
\author{Ngoc Thang}{Vu}

\affiliation[nocounter]{Institute for Natural Language Processing}{University of Stuttgart}{Germany}
\email{sarina.meyer@ims.uni-stuttgart.de}

\usepackage{comment}
\usepackage{cite}
\usepackage{subcaption}
\begin{document}

\maketitle

\begin{abstract}
    The performance of a voice anonymization system is typically measured according to its ability to hide the speaker's identity and keep the data's utility for downstream tasks. This means that the requirements the anonymization should fulfill depend on the context in which it is used and may differ greatly between use cases. However, these use cases are rarely specified in research papers. In this paper, we study the implications of use case-specific requirements on the design of voice anonymization methods. We perform an extensive literature analysis and user study to collect possible use cases and to understand the expectations of the general public towards such tools. Based on these studies, we propose the first taxonomy of use cases for voice anonymization, and derive a set of requirements and design criteria for method development and evaluation. Using this scheme, we propose to focus more on use case-oriented research and development of voice anonymization systems.
\end{abstract}

\section{Introduction}
The goal of voice anonymization is to modify a speech recording in such a way that it does not contain any information that allow to recognize the speakers in it. At the same time, it should still be possible to use the audio for other purposes. These could be the interaction with a voice assistant \cite{champion_2020_a-study, mawalim_2021_improving, zhang_2024_privacy}, a therapy session \cite{franzreb_2024_towards} or the training of a text-to-speech (TTS) system \cite{huang_2024_multi-speaker}. As these downstream applications have very different requirements to an audio, the exact setup of a voice anonymization system depends on its use case.

Most approaches follow the definitions of the Voice Privacy Challenge (VPC) which has been held in biennial rhythm since 2020 \cite{tomashenko_2022_vpc2020}. In this challenge, the downstream tasks are specified by models used for objective evaluation and by subjective annotations. The exact evaluation setup and thus the utility focus changes each VPC edition which also affects the focus in the research community. For example, in VPC 2020, utility was evaluated via speech recognition (ASR) and subjective naturalness and intelligibility, whereas in 2024 \cite{tomashenko_2024_vpc2024}, it was ASR and speech emotion recognition (SER) without subjective evaluation.

While it is known in the research community that the downstream task affects the structure and requirements of the voice anonymization system, in most publications, the downstream task remains underspecified and needs to be deduced from the evaluation metrics. The task is often framed as a downstream machine learning model that is applied to the anonymized data for example for ASR \cite{tomashenko_2022_vpc2020, tomashenko_2022_vpc2022, tomashenko_2024_vpc2024}, SER \cite{nourtel_2021_evaluation,zhang_2023_voicepm , tomashenko_2024_vpc2024}, or health state detection \cite{ghosh_2024_anonymising, ravuri_2022_preserving, arasteh_2024_addressing, zhu_2024_on}. However, voice anonymization is not only relevant if the audio is used for an automatic prediction method but also in situations where no such model is applied, such as online conversations between strangers. In such use cases, the requirements for voice anonymization systems differ from the one for downstream model applications. 

\begin{figure}
    \centering
    \resizebox{\columnwidth}{!}{
    \includegraphics[scale=1.0]{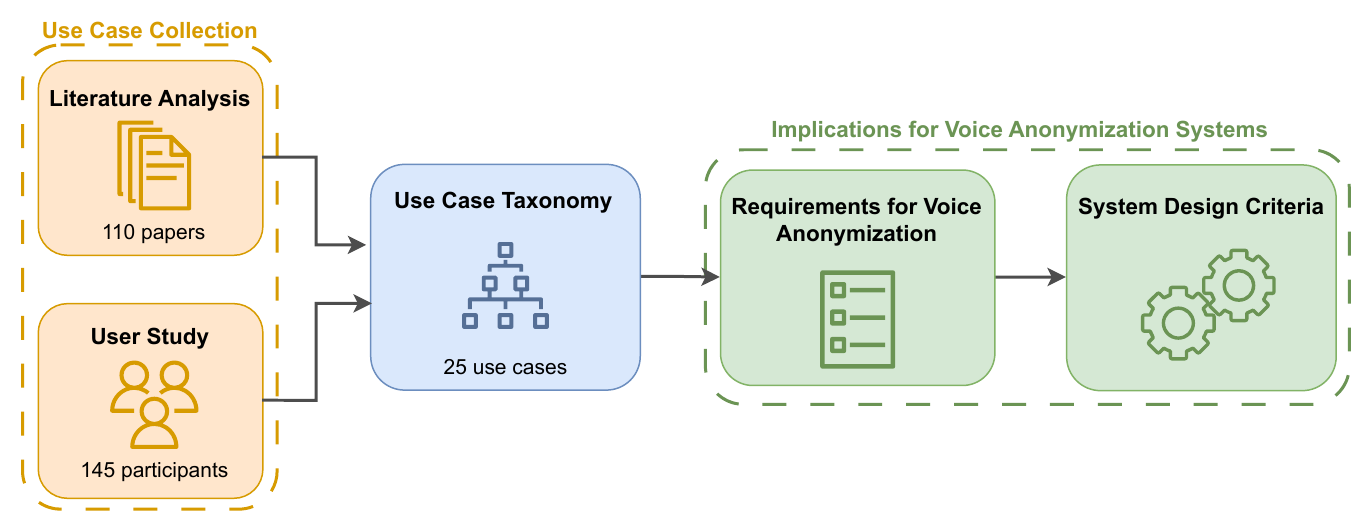}
    }
    \caption{Methodology of this paper. We perform a use case collection consisting of a literature analysis and user study to develop a use case taxonomy for voice anonymization. Using this categorization, we define use case-dependent requirements and design criteria for anonymization systems.}
    \label{fig:method}
\end{figure}

Although the need for a taxonomy of use cases has been expressed before \cite{nautsch_2019_gdpr}, to our knowledge, there has not been an extensive discussion of possible use cases for voice anonymization and their implications for the design of such systems so far. In this paper, we thus close this gap by creating a first taxonomy of voice anonymization use cases. As shown in Figure \ref{fig:method}, we base this taxonomy on an extensive use case collection consisting of a literature analysis and a user study. We analyze how different use cases affect the requirements an anonymization method needs to fulfill, and define system design criteria based on these requirements. 

Our contributions are the following:
\begin{itemize}
    \item We propose a taxonomy of use cases for voice anonymization based on expert and non-expert perspectives and expectations. We collect these opinions using an extensive literature review and a large-scale international user study.
    \item We define requirements that an anonymization needs to fulfill in different use cases and that influence the design of voice anonymization approaches.
    \item Based on these requirements, we propose a use case-based development scheme consisting of a list of design criteria that help to decide how to develop and evaluate an anonymization method for a specific use case.
\end{itemize}

\section{Background and Related Work}

\subsection{Voice Anonymization as defined in the VPC} \label{subsec:rw_anon}

In the speech research community, the term \textit{Voice Anonymization} (speaker anonymization, de-identification) refers to the task of manipulating a speech audio such that the identity of a speaker cannot be detected anymore while the audio is still usable for a specific purpose, generally called the \textit{downstream task}. In 2020, after a few individual methods have been proposed previously\cite{jin_2009_speaker, jin_2009_voice, pobar_2014_online, abou-zleikha_2015_a-discriminative, justin_2015_speaker, magarinos_2017_reversible, qian_2017_voicemask, bahmaninezhad18_odyssey, qian_2018_towards, fang_2019_speaker}, the first common definitions and standards for the task were defined by the first VPC \cite{tomashenko_2020_introducing, tomashenko_2022_vpc2020}. Since then, most publications follow these definitions or the versions of the VPC 2022 \cite{tomashenko_2022_vpc2022} or VPC 2024 \cite{tomashenko_2024_vpc2024}. 

In the VPC 2020 \cite{tomashenko_2020_introducing, tomashenko_2022_vpc2020}, the task is described as a game between users publishing data and attackers aiming to extract private information from this data. To prevent attackers from achieving this, users seek to remove personal, identifying information from the data while ensuring that all downstream goals can be achieved. For this, a voice anonymization method is designed that should (a) return a speech waveform, (b) hide the identity of the speaker, (c) not change other speech characteristics, (d) keep the same pseudo-voice for all utterances of a speaker, with different pseudo-voices for different speakers. Condition (c) was changed to linguistic content and paralinguistic attributes in VPC 2022 \cite{tomashenko_2022_vpc2022}, and to linguistic and emotional content in VPC 2024 \cite{tomashenko_2024_vpc2024}. The latter further dropped condition (d). While the exact nature of the downstream goals is not specified in the VPC descriptions, (c) is evaluated using a range of objective and subjective utility metrics, such as ASR, pitch correlation, SER, or human annotations of naturalness and intelligibility. The privacy condition (b) has been further evaluated with different automatic speaker verification (ASV) methods.

These definitions lack a certain specificity, especially regarding condition (c). It is not clear if utility should only be preserved for the specific models used during evaluation, or if it is expected to e.g., keep all paralinguistic attributes that do not correspond to speaker identity. In the case of the latter, it is not clear how to separate identifying attributes from non-identifying ones. These vague conditions lead to a mismatch in expectations and assumptions between groups. For example, it has been argued that a cascading system of ASR and TTS is not suitable for voice anonymization \cite{shamsabadi_2023_differentially, panariello_2024_the-voiceprivacy}, yet, this is not reflected in the task definitions and thus such systems have been proposed by other researchers \cite{meyer_2022_cascade, lee_2024_voice}. Without specifying a use case for the anonymization, it is difficult to come to a consensus about what approaches are suitable or even acceptable.

\subsection{Use Cases in Voice Anonymization}
The topic of use cases has been addressed briefly in previous voice privacy research. For instance, \cite{bäckström_2024_privacy} states that the main applications for speech technology are telecommunication and human-computer interfaces, and discusses privacy for these cases. However, the work stays only on a high level in summarizing what kind of sensitive information is processed in a specific application type (e.g., ASR), and how its processing influences privacy. 
By examining the legal and technical situation of privacy in speech data, \cite{nautsch_2019_gdpr} describe how and when different data sensitivity categories apply to speech data. They conclude that there is a lack of common understanding between legal and technical communities and thus a need for taxonomies across several dimensions, including use cases. To our knowledge, such a taxonomy has not yet been proposed. 
\cite{rahman_2024_scenario}, however, identified a lack of detail and consistency in papers presenting voice anonymization approaches. They therefore propose a scenario of use scheme to specify the attack and protection models an approach has been designed for. While this scheme would certainly facilitate the comparison of different approaches, it does not support researchers in their decision on how to design a system.
On the other hand, \cite{williams_2022_new} examine techniques and challenges of content privacy in audio recordings. They identify a need in specifying the downstream application when choosing a masking technique because different factors influence the applicability of and requirement for the privacy-preserving method. While focusing on private information in speech content rather than voice characteristics, their categorization of privacy threats according to the way how data is transmitted between different kinds of human or machine actors and how it can be intercepted resembles our use case taxonomy in Section \ref{sec:use_cases}.

As addressed in Section \ref{subsec:rw_anon} and further discussed in Section \ref{subsec:lit_review}, most voice anonymization publications do not specify what kind of use case they target. A few works, however, approach a specific scenario and evaluate their anonymization for that, such as psychotherapy sessions \cite{franzreb_2024_towards}, civic dialogue networks \cite{kang_2024_anonymization}, various health diagnostic settings \cite{ghosh_2024_anonymising, ravuri_2022_preserving, arasteh_2024_addressing, zhu_2024_on}, and the anonymization of training data for machine learning models \cite{huang_2024_multi-speaker, miao_2024_synvox2}. These approaches are still preliminary work for a few specific use cases but they already show that there is a wide variety of different recording settings, target groups, and general objectives. In this paper, we will address these issues in more detail.

\section{General Application Scheme of Voice Anonymization} \label{subsec:roles}
\begin{figure}
    \centering
    \resizebox{\columnwidth}{!}{
    \includegraphics[scale=1.0]{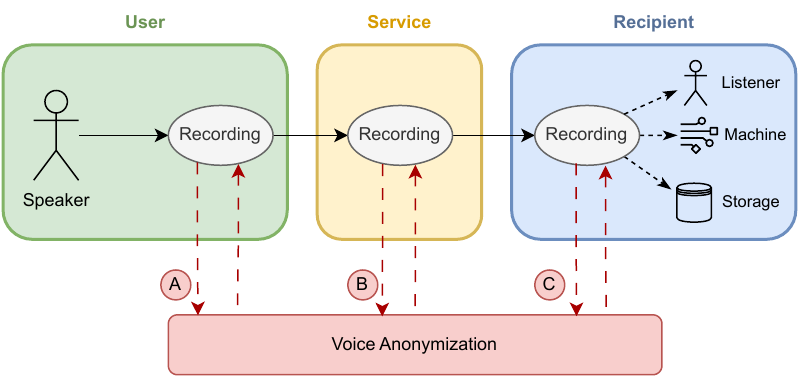}
    }
    \caption{Process of a recording in a general voice anonymization application and actors involved in it. There are three options where to perform voice anonymization: (a) on the user side, (b) on the service side, (c) on the recipient side.}
    \label{fig:roles}
\end{figure}

In a general sense, a \textit{use case} or \textit{application} for voice anonymization is a situation involving audio containing human speech (\textit{speech recording}), a reason for why this recording was created and/or is shared (\textit{purpose}), and one or several reasons for why it should be anonymized (\textit{threats}). Within such a situation, there are different actors interacting with the recording, with different interests regarding its contents and modifications. A simplified scheme of these actors is shown in Figure \ref{fig:roles}. 

The first actor is the \textit{speaker} whose speech is being recorded and processed. We will use the terms \textit{speaker} and \textit{user} interchangeably for this role, even if the speaker is not aware of being recorded. Regardless of whether the speaker knows of or has access to the recording, the first version of it exists on the user side before then being given to some \textit{service}. This service might just transmit the recording to its final destination, e.g., as an online meeting platform. It might also perform some simple processing steps or store the audio for a while until it can be further processed by the last actor, the \textit{recipient}. The recipient is closely linked to the purpose of the recording, and can be one or several humans (\textit{listener}), algorithms and models processing the audio (\textit{machine}), or a long-term \textit{storage} of the recording. Some actors are combined as the same entity, e.g., the transmission service might be the same as the recipient of the recording.

Each actor has their own version of the recording. Voice anonymization could be performed at any of those stages (A-C in Fig. \ref{fig:roles}), depending on the use case and factors like trust into each actor, security of the transmission, and device limitations. It is commonly assumed that the anonymization should either be performed on the user (A) or service side (B), due to a lack of trust in the recipient or the requirement to not have to change the downstream application. However, there could be a scenario in which the anonymization on the side of the recipient (C) is preferred, for example when creating a speech dataset. 

When developing a voice anonymization method, it is necessary to consider at which point the anonymization should be performed. This affects technical restrictions but also which party can influence the anonymization because each actor has different interests in the recording. The user might want the recording to be only usable for one specific purpose while the recipient might have an interest in using it for several different tasks. 
Thus, in the one case, anonymization should remove as much information as possible and keep only what is necessary for the downstream task, while in the other case, as much information as possible should be kept and only identifying information should be removed. 
Which of these strategies to follow, must be decided at an early stage of the system development.

\section{Use Case Collection}
Since no extensive examination of use cases for voice anonymization has been performed yet, we first collect statements about applications of such systems from voice privacy literature. This literature analysis gives insights into what kind of scenarios and objectives research has been targeting so far, and how the success of voice anonymization in these scenarios has been evaluated. As the literature only mirrors the perspective of experts on this task, we extend this analysis by performing a large-scale user study. By including answers of members of the general public across the globe, we aim to understand when and why non-expert users of speech technology would want to have voice anonymization being performed.
We analyze and categorize the different statements about anonymization use cases, motivations, and requirements from literature and user study using techniques of qualitative content analysis \cite{hsieh_2005_three}. We start by collecting all statements for a topic from a subset of publications or responses, group them into categories according to their similarity, and assign the remaining statements to these categories. New categories are formed if necessary and the overall categorization is checked and revised at the end.

\subsection{Literature Analysis} \label{subsec:lit_review}
In order to understand the research perspective on voice anonymization, we analyze a collection of 110 publications \cite{jin_2009_speaker, jin_2009_voice, pobar_2014_online, abou-zleikha_2015_a-discriminative, justin_2015_speaker, magarinos_2017_reversible, qian_2017_voicemask, bahmaninezhad18_odyssey, qian_2018_towards, fang_2019_speaker, aloufi_2020_privacy-preserving, champion_2020_a-study, han_2020_voice-indistinguishability, srivastava_2020_design, yoo_2020_speaker, kai_2021_lightweight, mawalim_2021_improving, patino_2021_speaker, prajapati_2021_voice, qian_2021_speech,stoidis_2021_protecting, agarwal_2022_speaker, champion_2022_are, chang_2022_zero-shot, chen_2022_privacy-utility, costante_2022_using, dubagunta_2022_adjustable, hernandez_2022_self, hu_2022_speechhide, maouche_2022_enhancing, mawalim_2022_speaker, meyer_2022_cascade, meyer_2022_speaker, miao_2022_analyzing, miao_2022_language, oreilly_2022_voiceblock, perero-codosero_2022_x-vector, ravuri_2022_preserving, tavi_2022_improving, turner_2022_generating, yuan_2022_deid-vc, chen_2023_voicecloak, deng_2023_vcloak, gupta_2023_voice, lv_2023_salt, meyer_2023_anonymizing, meyer_2023_prosody, miao_2023_speaker, nespoli_2023_two-stage, shamsabadi_2023_differentially, tran_2023_a-speech, yao_2023_distinguishable, akti_2024_voice, blouch_2024_tuning, cai_2024_privacy, chen_2024_an, chen_2024_modeling, chen_2024_reprogramming, cheng_2024_saic, das_2024_comparing, das_2024_speecher, ghosh_2024_anonymising, gu_2024_a-voice, huang_2024_diffvc, lee_2024_voice, matassoni_2024_speaker, meyer_2024_probing, panariello_2024_preserving, panariello_2024_speaker, pohlhausen_2024_enhancing, quamer_2024_end-to-end, singh_2024_voice, wang_2024_asynchronous, webber_2024_voice, yao_2024_distinctive, chen_2025_pseudo, miao_2025_a-benchmark, yao_2025_musa, maouche_2020_a-comparative, nautsch_2020_zebra,noe_2020_noe, srivastava_2020_evaluating, tomashenko_2020_vpc_results, champion_2021_evaluating, champion_2021_on, nourtel_2021_evaluation, baeg_2022_dnn, liu_2022_a-novel, franzreb_2023_a-comprehensive, panariello_2023_vocoder, zhang_2023_voicepm, arasteh_2024_addressing, franzreb_2024_towards, huang_2024_multi-speaker, leschanowsky_2024_voice, meyer_2024_voicepat, panariello_2024_the-voiceprivacy, williams_2024_anonymizing, zhang_2024_privacy, zhu_2024_on, tomashenko_2025_analysis, nautsch_2019_preserving, tomashenko_2020_introducing, obrien_2021_anonymous, tomashenko_2022_vpc2020, tomashenko_2022_vpc2022, kang_2024_anonymization, miao_2024_synvox2, rahman_2024_scenario, tomashenko_2024_vpc2024}.
Most papers present anonymization approaches \cite{jin_2009_speaker, jin_2009_voice, pobar_2014_online, abou-zleikha_2015_a-discriminative, justin_2015_speaker, magarinos_2017_reversible, qian_2017_voicemask, bahmaninezhad18_odyssey, qian_2018_towards, fang_2019_speaker, aloufi_2020_privacy-preserving, champion_2020_a-study, han_2020_voice-indistinguishability, srivastava_2020_design, yoo_2020_speaker, kai_2021_lightweight, mawalim_2021_improving, patino_2021_speaker, prajapati_2021_voice, qian_2021_speech,stoidis_2021_protecting, agarwal_2022_speaker, champion_2022_are, chang_2022_zero-shot, chen_2022_privacy-utility, costante_2022_using, dubagunta_2022_adjustable, hernandez_2022_self, hu_2022_speechhide, maouche_2022_enhancing, mawalim_2022_speaker, meyer_2022_cascade, meyer_2022_speaker, miao_2022_analyzing, miao_2022_language, oreilly_2022_voiceblock, perero-codosero_2022_x-vector, ravuri_2022_preserving, tavi_2022_improving, turner_2022_generating, yuan_2022_deid-vc, chen_2023_voicecloak, deng_2023_vcloak, gupta_2023_voice, lv_2023_salt, meyer_2023_anonymizing, meyer_2023_prosody, miao_2023_speaker, nespoli_2023_two-stage, shamsabadi_2023_differentially, tran_2023_a-speech, yao_2023_distinguishable, akti_2024_voice, blouch_2024_tuning, cai_2024_privacy, chen_2024_an, chen_2024_modeling, chen_2024_reprogramming, cheng_2024_saic, das_2024_comparing, das_2024_speecher, ghosh_2024_anonymising, gu_2024_a-voice, huang_2024_diffvc, lee_2024_voice, matassoni_2024_speaker, meyer_2024_probing, panariello_2024_preserving, panariello_2024_speaker, pohlhausen_2024_enhancing, quamer_2024_end-to-end, singh_2024_voice, wang_2024_asynchronous, webber_2024_voice, yao_2024_distinctive, chen_2025_pseudo, miao_2025_a-benchmark, yao_2025_musa} but we also include work focusing on evaluation techniques for voice anonymization \cite{maouche_2020_a-comparative, nautsch_2020_zebra,noe_2020_noe, srivastava_2020_evaluating, tomashenko_2020_vpc_results, champion_2021_evaluating, champion_2021_on, nourtel_2021_evaluation, baeg_2022_dnn, liu_2022_a-novel, franzreb_2023_a-comprehensive, panariello_2023_vocoder, zhang_2023_voicepm, arasteh_2024_addressing, franzreb_2024_towards, huang_2024_multi-speaker, leschanowsky_2024_voice, meyer_2024_voicepat, panariello_2024_the-voiceprivacy, williams_2024_anonymizing, zhang_2024_privacy, zhu_2024_on, tomashenko_2025_analysis}, papers describing the VPC \cite{tomashenko_2020_introducing, tomashenko_2022_vpc2020, tomashenko_2022_vpc2022, tomashenko_2024_vpc2024} and other papers about voice anonymization \cite{nautsch_2019_preserving,  obrien_2021_anonymous, kang_2024_anonymization, miao_2024_synvox2, rahman_2024_scenario}. We analyze these papers according to three aspects: (a) the anonymization situation (use case) and reason (privacy threat), (b) the requirements for the anonymization (i.e., which information to keep and which to remove during the process), and (c) the evaluation that is either performed or proposed for an anonymization system. We give the percentage of papers that mention a certain aspect to estimate its relevance in the research community.

\subsubsection{Anonymization Situation} \label{subsubsec:lit_use_cases}
Most papers (72\%) mention one or several general use cases for voice anonymization. 
Half of the papers (51\%) highlight the automatic processing of speech data, e.g., to preserve user privacy during their interactions with voice assistants 
and IoT, tools analyzing the medical conditions of users, or for using speech as training data in machine learning. 35\% of papers see a benefit in anonymizing data that is intended for human listeners, such as in social media, medical consultations, customer service, or court recordings. Another type of use cases includes the general anonymization in data collections, mentioned in 22\% of papers. 

Although most papers state such use cases as motivations, only 
16\% explicitly express which use case they are aiming for, whereas the majority leaves this open. If it is specified explicitly, the approaches are mainly designed for health applications, anonymizing model training data, or data sharing.

53\% of papers also give one or several reasons for why voice anonymization should be performed. These reasons are generally divided into three attack categories. The main one is protection against spoofing attacks (35\%), usually in the form of voice cloning to get access to personal information or to generate harmful fake content about a speaker. Another threat are profiling attacks (26\%) in which personal information about a user (e.g., their age) is extracted, often from different sources, in order to create a profile of that user, e.g., for targeted advertisement. Finally, linkage attacks (20\%) describe a scenario in which data is linked to a specific person, usually by using speaker recognition.

\subsubsection{Anonymization Requirements}
The requirements of an anonymization system are usually specified along two dimensions: what information the system should remove (i.e., anonymize) and what it should keep (i.e., preserve) from the input audio. Naturally, the main type of information that is aimed to be removed is anything related to speaker identity (94\%). However, also other personal attributes could be aimed to be removed, such as gender (4\%), emotion (1\%), or sensitive speech content (4\%). 

There is less consistency when it comes to what information needs to be kept. 
The preservation of linguistic information is often seen as a natural requirement of voice anonymization \cite{panariello_2024_the-voiceprivacy} but is only named in 65\% of papers. Several papers state that the intelligibility should be preserved (18\%). 
Although similar, it is important to note that intelligibility and the preservation of linguistic content are distinct concepts. 
An anonymized audio might be intelligible but of different linguistic information than the original, or it might degrade the rate of intelligibility while keeping the exact words. Other information that is regularly targeted to be preserved are naturalness (24\%), emotions (14\%), voice or audio quality (11\%), voice distinctiveness (10\%), and prosody (10\%). 4 papers \cite{agarwal_2022_speaker, chen_2022_privacy-utility, oreilly_2022_voiceblock, wang_2024_asynchronous} aim for \textit{asynchronous anonymization} in which identity information is only removed for machine inference but unchanged for human perception, e.g., by adding adversarial noise to the audio.

\subsubsection{Evaluation}
Almost all papers present some form of objective evaluation (96\%) while a subjective evaluation is performed in 33\% of cases. 
The objective evaluation is often split into assessing the privacy requirements (what is removed) and utility requirements (what is kept). For privacy evaluation, speaker verification (76\%) and identification (18\%) are most often performed. Utility evaluation often includes the use of ASR (75\%), SER (15\%), or metrics to assess voice distinctiveness (16\%), pitch correlation (13\%), and audio or voice quality (15\%). Subjective evaluation is often measured as naturalness (17\%), intelligibility (12\%), speaker similarity (13\%) or audio quality (10\%).

We observed that not all papers are consistent in evaluating all requirements that they mention in a paper. For example, 12 papers (11\%) claim to keep naturalness during anonymization but do not evaluate this.

\subsection{User Study}
The user study was performed on Prolific\footnote{\url{https://www.prolific.com}} from September 2024 to May 2025. Besides demographic information, participants were asked to answer in free text form in what scenarios and why they could imagine voice anonymization should be used, both in \textit{general} and specifically for their \textit{personal} voice. Participation in the study took around 5-7 minutes and was compensated according to minimum wage in Germany (12.80 Euro per hour). In total, 170 participants from around the world who were fluent in English were recruited. 25 had to be excluded due to off-topic answers, resulting in a total of 145 participants. The pool of participants was diverse in terms of gender (52\% female, 46\% male, 2\% other or unknown), and age (48\% below 30 years, 28\% between 30 and 39, 13\% between 40 and 49, 10\% above 50 years). The participants came from a set of 24 countries (40\% from Europe, 26\% from Africa, 29\% from North America, 6\% other), with the majority coming from the US (28\%), South Africa (24\%), and United Kingdom (23\%). 74\% of participants stated that English was (one of) their native language(s), with 15\% of participants giving multiple native languages. Most participants had a higher education (44\% Bachelor's, 24\% Master's, 6\% PhD as highest degree), and all participants stated to have at least some basic knowledge about artificial intelligence.

\subsubsection{Anonymization Use Cases} \label{subsubsec:user_study_use_cases}
Almost all participants named at least one use case for \textit{general} and \textit{personal} scenarios. A few participants (9\%) stated that they would not want their voice to be anonymized, either because they do not trust the technology or because they would find it confusing. Otherwise, there were no clear differences in answers for \textit{general} and \textit{personal} situations, so we merge the answers to both questions together and give the percentages of participants naming a use case for either question.

The use case that has been mentioned most involves \textit{legal situations}, e.g., when a victim or witness gives a statement in court (33\%). Other frequently named situations are \textit{online interactions with strangers} (20\%) and \textit{calls to customer service} (18\%). Given that the participants are likely to frequently take part in other studies on Prolific, 19\% also mentioned \textit{research situations} involving speech.

Overall, a large variety of use cases were named which will be discussed in more detail in Section \ref{sec:use_cases}. Most participants (83\%) listed at least one use case concerning the interaction between humans. Human-computer communication and data storage scenarios were mentioned by 17\% and 32\%, respectively.

\subsubsection{Reasons for Anonymization}
80\% of participants gave not only the use cases but also reasons for why the voice should be anonymized in these scenarios. The main reason is the wish to simply \textit{hide the identity or stay anonymous} (51\%), without further explanation of why this should be necessary. Furthermore, \textit{safety} (27\%), like protection from retaliation, and protection from \textit{harassment} (6\%) were given by several participants. Other important reasons are to ensure the \textit{confidentiality} of shared information (12\%), especially when talking about medical or financial topics, and to \textit{protect the voice from being cloned} or used in \textit{model training} (9\%). 8\% of participants wish to increase the \textit{security} of their data through anonymization, e.g., in case of data leaks, and 6\% want to avoid a \textit{misuse} of their data without consent. 

Besides protecting their identity, several participants also mentioned reasons that involved hiding certain attributes in their voice, for example because they \textit{disliked their voice} (3\%), want to generally \textit{hide how it sounds like} (3\%), or do not want to \textit{reveal certain aspects} about them like their gender or accent (7\%). 10\% of participants further stated that voice anonymization would give them \textit{confidence to speak freely} and give candid responses. This shows that voice anonymization should not only be seen as a technique for privacy protection, but also to reduce bias, fear, and lack of confidence.

\subsection{Comparison of Literature and User Study}

When comparing the summaries of use cases described in literature and user study in Sections \ref{subsubsec:lit_use_cases} and \ref{subsubsec:user_study_use_cases}, it is clear that there are certain differences in what researchers focus on when developing a voice anonymization system and what users expect from it. While use cases involving the interaction between humans and computers are given in 51\% of papers and most papers rely only on objective instead of subjective evaluation, such use cases are only mentioned by 17\% of participants in the user study. On the other hand, 83\% of participants list use cases for human-human interactions, which are only named in 35\% of papers. This suggests that there are needs for voice anonymization from the perspective of the general public that are not addressed in research. We also observed a larger diversity of use cases and reasons for anonymization given by participants than in publications. While the literature analysis might not fully reflect all assumptions and motivations of researchers but only the ones explicitly stated in papers, it indicates that researchers might need to broaden their perspective on possible applications for voice anonymization.

\section{Use Case Taxonomy} \label{sec:use_cases}

\begin{figure}
    \centering
    \resizebox{\columnwidth}{!}{
    \includegraphics[scale=1.0]{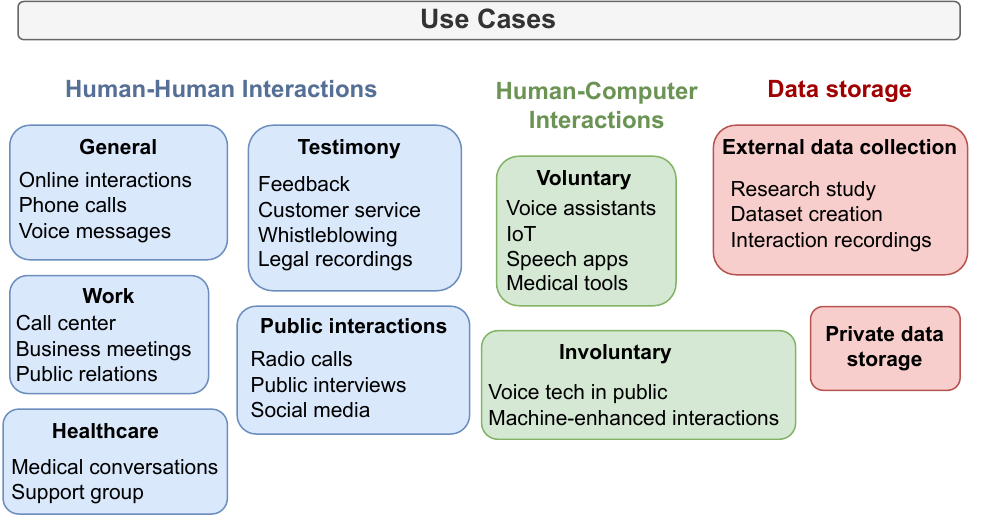}
    }
    \caption{Taxonomy of use cases for voice anonymization.}
    \label{fig:use_cases}
\end{figure}

Based on the use cases named in the literature and user study, we create 25 use case categories by grouping the single use cases based on the purpose of the recording, the nature of its contents, or the voluntariness of its creation. Following the distinction between recipients described in Section \ref{subsec:roles}, we divide the categories into three main groups (\textbf{human-human interactions}, \textbf{human-computer interactions}, \textbf{data storage}) and 9 subgroups. The overall taxonomy is shown in  Figure \ref{fig:use_cases}. 

Note that some recordings can be used for several use cases, e.g., the interaction with a voice assistant and the usage of that data for model training. Since the categories are purely based on the outcome of our use case collection, there might be use cases that are not included here. Furthermore, some participants in the user study stated that anonymization should always be possible for any kind of recordings, either to have it as a personal choice, or because certain user groups like children or people in witness protection programs should always be protected. 

\subsection{A: Human-Human Interactions}
This category comprises situations in which speech data is recorded and shared for a purpose in which people listen to this data, mostly as a means of communication. The recipients might or might not be known to the user, and could be one person or a group of people. 

\textbf{A1: General.} These are use cases without a very specific purpose but just with a general communicative aim. This includes \textbf{online interactions} where users want to keep the anonymity of their online presence also in voice conversations. Similarly, in \textbf{phone calls}, especially when the number of the caller is suppressed or the caller is a stranger, users would like to have the option to keep their voice anonymous in case the other person is a scammer trying to record the user for future voice cloning scams, or in other way untrusted. A third use case in this group is \textbf{voice messages}. Some users are aware that these voice messages would be saved on servers that might not be secure enough and therefore want to keep their voice identity hidden. Others stated that they might not know or trust the receiver of the voice message who could for example forward the message to other people without the user's consent.

\textbf{A2: Testimony.} This subcategory contains use cases in which the user is giving some kind of testimony. The use cases differ in their gravity and impact, and range from general feedback situations to crime revealings. As such, the lowest stakes are in situations where users give \textbf{feedback} in an environment in which they are known as a person. This could be at the work place or in any other situation in which the user is known to the receivers of the feedback, and wishes to stay anonymous in order to be able to freely express their concerns or complaints. A second scenario is the classic \textbf{customer service} situation. Here, it is unlikely that the user is known personally to the customer service employee but needs to be aware that other people might listen to the call or might even record it. Depending on the kind of service, serious complaints might result in disadvantages for the user. 
Anonymity is crucial if the testimony reveals crimes or ethical misconduct, and the user needs to be protected from retaliation. This is the case for \textbf{whistleblowing} in which an insider reveals secret information about an organization. Another situation are \textbf{legal recordings} which could be statements by witnesses or victims about a crime. The main difference between whistleblowing and legal recordings is that whistleblowers generally aim to reach the general public with their information, while legal recordings are usually intended for a smaller group of people such as the police or in court. 

\textbf{A3: Privacy at work.} This category consists of scenarios in which the user is at work and therefore might need to do or say something that they do not want to be connected to their personal identity. This could be in a \textbf{call center} situation in which the call center employee should be anonymous such that the caller is not aware of their identity in case they want to express their dissatisfaction with a company in the form of personal retaliation. % at an employee. 
It could also be in \textbf{business meetings}, with company internal or external members, in which candid comments might be discouraged unless voices are anonymized. It could also be a work-related recording that is directed to the general public in the form of a \textbf{public relations} task. Especially if the job is controversial or even dangerous, it is important to give the option to hide the speaker's identity.  Even if the user themselves agrees with the values of their own work, they might live in an environment where it would be dangerous to admit to this.

\textbf{A4: Public interactions.} Related to the previous use case, any recording that is intended for a large group of unknown people can pose a certain privacy risk. A user cannot be certain that the recording would not be listened to by someone they know personally, or that the data will not be misused. Especially if the contents of the recording are sensitive, for example in case of \textbf{public interviews} or talks about specific topics that are broadcasted to TV, radio or online platforms, anonymization might be important. This could also apply to other public interactions, such as \textbf{radio calls} in which the user calls a radio station (e.g., to participate in a quiz) and this call gets broadcasted. In modern times, online presence in \textbf{social media} is more prevalent. Subjects stated that they would like to have their voice anonymized in social media such that their audio cannot be used for threats like voice cloning, that their personal contacts cannot identify them, or that their voice cannot be used as background track in other social media posts.

\textbf{A5: Healthcare.} Any situation that deals with sensitive information about a person making them vulnerable requires special protection. This is especially true in case of healthcare topics. One scenario are \textbf{medical conversations} such as online medical consultations or therapy sessions. Another use case are situations in which a person seeks for help, for example in a \textbf{support group} or helpline setting. Offering the option to anonymize the voice of such people or patients in general might encourage more people to seek help without having to fear personal shame or legal consequences.

\subsection{B: Human-Computer Interactions}
In human-computer interactions, speech recordings are intended to be processed by machines, for example to perform ASR or analysis tasks. 
We distinguish between voluntary and involuntary recordings in this category.

\textbf{B1: Voluntary Interactions.} Recordings in which the user chooses to interact with a machine via voice count as voluntary interactions. This includes \textbf{voice assistants}, \textbf{IoT}, and smaller \textbf{speech apps} that are designed for a specific purpose (e.g., transcription or language learning). 
These applications might process the data on device or in the cloud, and differ in the kind of information they process.
Another use case in this category are \textbf{medical tools} that are intended to support human practitioners in diagnosing and monitoring patients' health states either within or outside of medical sessions.

\textbf{B2: Involuntary Interactions.} This category consists of use cases in which a user does not get to choose whether they want to interact with a machine, and might not even be aware that their speech is being recorded and processed. These are generally situations of \textbf{voice tech in public} places. This comprises surveillance technology but also situations where there was never an intention of processing the data of uninformed people, such as voice assistants in cafes or doctor's offices. In such cases, it would be necessary to detect and filter out which interaction is intentional and should be kept, and which should be discarded or at least anonymized. Another case are situations that we refer to as \textbf{machine-enhanced interactions}. These are typically interactions that are intended for a human recipient but automatically enhanced with voice technology, even if the user does not want this, e.g., phone calls that include an automatic processing of the request before transferring them to a human. Often, the user is forced to interact with the machine if they want to continue the call and might not know how their speech data is processed or stored.

\subsection{C: Data storage}
The third group consists of use cases of general data collection and storage. 
There are no immediate goals that the user tries to achieve with their recording but rather long-term storage or future processing purposes. The data might later be used in a use case from \textbf{A} or \textbf{B}, but this future application might not be defined at the time of recording and storage. Thus, data in this group generally require a broader utility than in the other cases.

\textbf{C1: External data collection.} These are use cases in which the speech data is recorded or shared with an external party, for example to create a voice dataset. This includes \textbf{research study} situations such as voice-based surveys and interviews, or \textbf{dataset creation} scenarios in which the data is intended for further analysis or model training. Another type are \textbf{interaction recordings} which are intended for a different use case but then stored by a company for quality improvement or training purposes, such as recorded customer service calls.

\textbf{C2: Private data storage.} There are also cases in which users might want to let external entities store their data but not process or access it. This would for example occur in case of voice messages or private videos stored in the cloud. Though this could also be solved by encryption, anonymization might be a viable option if the users do not care or even wish that the data will be altered.

\section{Implications for Voice Anonymization Systems}
\begin{figure}[]
    \centering
    \includegraphics[scale=0.4]{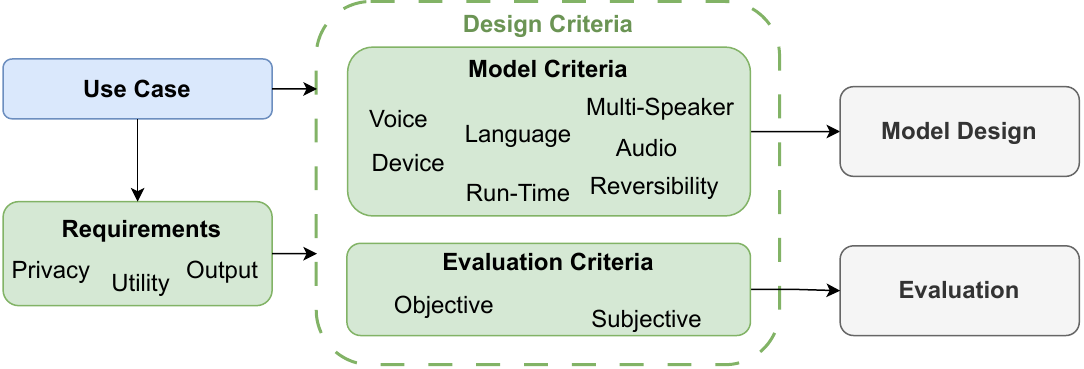}
    \caption{Requirements and design criteria for voice anonymization systems that are influenced by its use case.}
    \label{fig:design_criteria}
\end{figure}

The large number of diverse use cases shows that voice anonymization is needed in a variety of situations and environments. These range from daily life scenarios that most people experience on a regular basis to less common but particularly sensitive circumstances such as whistleblowing. We also found that hiding one's identity is not always the only purpose of voice anonymization. Sometimes, the feeling of being anonymous or someone else, a protection against voice cloning, secure data storage, or the reduction of attribute-based biases can be more important than complete anonymity.
Thus, depending on the use case, there are different conditions that a voice anonymization system needs to fulfill. As shown in Figure \ref{fig:design_criteria}, these are first reflected in different general requirements to the anonymized data, and in turn influence how the model and evaluation need to be designed. As a consequence, the choice of a target use case leads to strong implications for the development of an anonymization system. 

\subsection{Requirements for Voice Anonymization}
Each anonymization method needs to fulfill certain requirements for its data to be considered as anonymous and suitable for further processing. These requirements affect the general applicability of a method in a specific scenario and are therefore defined by its use case.

\subsubsection{Privacy requirement}
The first requirement addresses the question of what information to remove from the recording in order to achieve the required level of privacy. Generally, this defaults to all features that could be used by a detection model or a human listener to recognize the identity of any individual in an audio (\textit{identifying information}). However, depending on the use case, it might be necessary to extend this to other personal attributes such as age, gender, accent, or health state. For example, when anonymizing a recording where potential listeners might know the original speaker personally and where the pool of possible speakers is limited (e.g., in feedback scenarios), the user's accent might reveal their identity even if the voice attributes are changed. It is therefore necessary to evaluate the exact privacy requirements for a specific use case, and possibly even for a specific user. 

\subsubsection{Utility requirement}
The question of what information should be kept during anonymization is closely related to the purpose of the recording and thus its utility. In some cases, only information strictly necessary for a use case needs to be kept. For example, if an audio is only intended for ASR (e.g., as part of a dictation app), the anonymization might not need to preserve the naturalness or voice diversity of the audio. 
On the other hand, especially in use cases involving human-human interaction, it might be preferable to keep more than what is strictly necessary. For example, in a phone call, not only the linguistic content should be preserved and intelligible, but the voice should also be rather pleasant to listen to. 
For some use cases, it might be beneficial to not keep the respective property as displayed in the original audio but to modify them in a way that they improve utility. 
For example, if the speaker has a strong accent that the recipient might struggle to understand, the anonymization could change the accent to a more familiar one. In a multi-speaker environment, increasing the perceptual differences between voices might make it easier for listeners to distinguish between different speakers.

\subsubsection{Output requirements}
One basic definition of voice anonymization is that the output should be a waveform. However, the exact specifications for the output format are not given. For example, it is rarely specified if the audio should have the same length and speed as the original, or if this could or should be changed. These decisions can affect the privacy protection \cite{tomashenko_2025_analysis} but also depend on the purpose of the recording or on other processing steps that are being performed in parallel to voice anonymization. For instance, the anonymized audio might need to be synced with a video that might itself have been anonymized \cite{franzreb_2024_towards}. It could also be that the linguistic content needs to be anonymized \cite{qian_2017_voicemask, williams_2023_exploratory} which might need to happen before, during, or after voice anonymization.

\subsection{System Design Criteria} \label{sec:design_criteria} 
Combined with the requirements, the use case further influences criteria of anonymization model design and its evaluation.

\subsubsection{Model Design Criteria}
The use case and requirements affect different criteria for model design, ranging from the design of the anonymization process to inference limitations.

\textbf{Target voice selection.} Since anonymization typically changes the appearance of the speaker's voice, the selection of the target voice is a big topic in the field. In some approaches, a reference speaker from a dataset is used as target speaker \cite{champion_2022_are, panariello_2024_speaker}, other works focus on creating voices that do not correspond to real speakers, either as a combination of a pool of reference speakers \cite{fang_2019_speaker, srivastava_2020_design, meyer_2022_speaker, miao_2022_language}, a modification of the original voice \cite{mawalim_2022_speaker, miao_2023_speaker, yao_2024_distinctive}, or by using a generation model \cite{turner_2022_generating, yuan_2022_deid-vc, meyer_2023_anonymizing, chen_2025_pseudo}. 
Besides impacting the performance of the anonymization, there are several aspects that should be considered when implementing a selection method:

\begin{itemize}
    \item \textbf{Voice Origin:} The first question is where the target voice should come from. Depending on the purpose, recipient and ethical factors, it might be acceptable or even beneficial to re-use or modify the voice of an existing speaker, or it might be necessary to generate an artificial voice. Selecting an existing speaker as the target voice is generally simpler but might be harmful for that speaker or confusing, especially if the voice is known to the recipient. 
    \item \textbf{Voice Diversity:} Another aspect is the diversity of different voices created by one anonymization system. In some cases, it might be acceptable if several users are anonymized with the same target voice, for example if the recordings are intended for different recipients. However, especially in situations involving several users, it might be necessary to select a new voice for each user or even each utterance. In those cases, it has to be decided how different these voices should be, and if this voice diversity should exactly match the diversity of the original voices (e.g., in a conversation).
    \item \textbf{Durability:} A more technical question is how long the assignment of original voice to target voice should be kept. This depends on if the target voice for one speaker should be consistent for several utterances, sessions or even all times the speaker uses the respective application. From a security and performance point of view, choosing a new voice for each utterance of a speaker is preferable because then speaker mappings do not need to be tracked or stored, and in multi-speaker recordings, speaker diarization only needs to separate utterances but not speakers. Furthermore, it is more stable against privacy attacks because the attacker cannot learn this mapping. However, in most use cases it would be distracting if the user's voice would change with each new utterance, for example in therapy sessions. Especially in recordings containing the voices of several speakers, the information about how many speakers and who speaks what could be lost. 
    \item \textbf{Voice Attributes:} Since a voice is connected to several attributes of a speaker, changing a voice would also affect how the user is perceived. In some cases, this is preferred, e.g., to reduce attribute-specific biases, but it might also be relevant to keep certain aspects, for example if they are given by the context or content. For instance, in a legal or medical recording, the gender of a speaker should generally be kept, whereas it might be preferable to choose features of an opposite or neutral gender for recordings at the work place. 
\end{itemize}

\textbf{Multi-speaker support.} The handling of multiple voices by different speakers in one recording has not been addressed much in voice anonymization research yet  \cite{miao_2025_a-benchmark}. However, several use cases consist of multi-speaker interactions, especially in cases of human-human conversations (e.g., online interactions, business meetings, support groups) and general data collections. If multiple voices are present in a recording, the first question is if all of them need to be anonymized or if some could or should stay unmodified (e.g., the therapist or interviewer). Another question is if all voices are combined in one audio channel, or if each voice comes from a different source (e.g., in online meetings). Depending on this, speaker diarization might need to be performed. Multi-speaker anonymization also affects the question of whether the anonymization or target voice selection should be performed on user- or service-side. If everything is performed on the user-side, it might be difficult to ensure sufficient voice diversity in the anonymized result.

\textbf{Language support.} While voice anonymization research often focuses on English-only data, several approaches for language-independent or multilingual anonymization have been proposed \cite{miao_2022_language, meyer_2024_probing, yao_2025_musa}. For a use case, it needs to be assessed which languages might be present in the audios and whether all of them need to be supported by the anonymization. Linguistic phenomena like code-switching and the use of foreign words should also be taken into account for this decision. 

\textbf{Run-time.} The run-time of different voice anonymization systems has been discussed in the context of designing methods that could be run in real-time or even streaming \cite{oreilly_2022_voiceblock, chen_2024_an, quamer_2024_end-to-end}. Such real-time anonymization systems are necessary for several use cases, such as phone calls, business meetings or support groups. In other use cases, a run-time that is a bit slower than real-time might be acceptable, such as the use of medical tools or social media. There are even some use cases in which anonymization could be slower and run in the background, e.g., the anonymization of datasets or legal recordings. One question related to the run-time is if the anonymized audio should be presented to the user for a check before it is sent to the recipient. Such a check could give the user the possibility to make sure that the anonymization worked correctly and produced audio of sufficient privacy, utility and quality. This is obviously not possible if the interaction is in real-time but could be useful for example for social media or feedback scenarios.

\textbf{Device limitations.} The device that the anonymization should run on restricts the space and resources that are available during the process. Thus, several lightweight approaches have been proposed \cite{kai_2021_lightweight, chen_2024_an, pohlhausen_2024_enhancing, quamer_2024_end-to-end}. The device limitations depend on whether the anonymization should be performed on the user-side or on a server. User-side often means that no GPU is available and that memory is limited. There is more freedom when performing the operation on a server, but this would require to transfer the non-anonymized audio to the server, posing additional privacy and security risks. A related question is whether an internet connection needs to be available during anonymization. These questions depend on the sensitivity of the speech contents and the severity of a privacy leak. 

\textbf{Audio characteristics.} A recording can include more than just voices but information related to the environment in which it has been created (e.g., background sounds, microphone type, room reverberation). If this is the case, it should be discussed how to process this. The information might give indications about the identity of the speaker and then needs to be changed during anonymization. In some situations or use cases, however, it might be preferred to keep certain aspects of the original audio characteristics. For example, if a person gives a statement to an interviewer during a protest, they might not want to remove or modify the background sounds as long as nobody could be identified from them. 

\textbf{Reversibility.} In the earlier works of voice anonymization, the ability to reverse an anonymization process to retrieve the original audio from it was seen as an requirement \cite{jin_2009_voice, abou-zleikha_2015_a-discriminative, magarinos_2017_reversible}. Nowadays, there is a consensus that anonymization should be irreversible instead \cite{nautsch_2019_preserving} to fully preserve speaker privacy. In fact, from a legal perspective, anonymization is defined as an irreversible process, while the reversible pendant is called pseudonymization \cite{kamocki_2022_pseudonymisation}. Especially in the case of private data storage, a reversible modification has the benefit that the user would be able to retrieve the data back in its near-original form. While reversible pseudonymization shares certain similarities to encryption, it might have some advantages over existing methods, and thus might be a preferred solution for a use case.

\subsubsection{Evaluation Design Criteria}
Another set of design criteria that depends both on the requirements and the model design concerns the evaluation of the anonymization system. Generally, everything that is defined by the requirements needs to be evaluated. Model design choices need to be included in the evaluation if they are important for a use case. For example, if the run-time is crucial for the suitability of an anonymization method in a certain scenario, it must be measured and reported.

Depending on the utility requirements, the utility evaluation can be either performed objectively, subjectively or both. Not every use case requires an objective and subjective utility evaluation. For example, if the anonymized recording is only intended for human recipients but not for machines, an objective utility evaluation such as ASR is not necessary and can only be used as an approximation of the human perception. On the other hand, if an audio should only be processed by a machine and not by a human, the naturalness of the audio likely does not matter and thus a subjective evaluation is not necessary.

Privacy evaluation should always be performed objectively to test the worst case scenarios for privacy. In most cases, privacy evaluation should measure the ability of the anonymization to hide speaker identity (e.g., using ASV) but it could be extended to other privacy metrics such that different privacy threats are considered, e.g., voice cloning. Privacy evaluation should include a subjective component if it is relevant that the anonymization can be perceived accordingly by human listeners. Besides empirical metrics, theoretical measures can be used to give mathematical proofs for privacy levels \cite{bäckström_2024_privacy}.

\section{Future Perspectives}
We are aware that the development and evaluation of voice anonymization systems is already complicated and time-consuming, and that even in simplified settings, the task is far from being considered as solved. By proposing to target specific use cases when developing a method, we do not aim to make the task more complicated, but instead to increase openness about assumptions and limitations underlying a technique. We do not expect researchers to fulfill all criteria for a use case or to perform the perfect evaluation. Instead, it is important to address the limitations of an approach or evaluation to raise the awareness that this method might not perform in all situations as presented. Furthermore, just because an approach does not achieve the perfect scores in privacy or utility evaluation, it is not automatically unsuitable for an actual application. Other factors like run-time, language support or perceptual quality might be more important for a use case than perfect anonymity, and even some imperfect anonymization might still be better than no protection at all. However, the (proven) performance and limitations should always be communicated to the users.

By developing methods regardless of a concrete use case, researchers tend to follow the evaluation framework of the VPC in order to achieve comparability to other methods. In this way, however, ideas that do not fit to this specific evaluation setup are being discouraged even though they might be beneficial in other anonymization scenarios. We therefore need to encourage researchers to explore different directions with different objectives within the realm of voice anonymization. To facilitate this, we propose to split this task into several subtasks, and to include separate tracks for each subtask in the VPC, e.g.: (a) Voice anonymization for human-computer interactions in which utility is defined by the performance of downstream models, (b) voice anonymization for either only human recipients or human and machine processing, in which factors regarding perceptual qualities of the audio need to be considered, (c) asynchronous anonymization \cite{wang_2024_asynchronous} in which the identifiability of speakers should differ between human perception and automatic recognition, and (d) (reversible) pseudonymization in which it is possible to reconstruct the original audio, in part or in full.

During our user study, we observed that there is a need to further educate the general public about the harm of generating false audio content (e.g., in the form of audio deepfakes) and about ethical uses of voice anonymization tools. Almost 10\% of participants stated that they would like to use voice anonymization technology to either improve the voice in their recordings or to prank other people. It is important to be aware that voice anonymization might be misused for illegal or unethical purposes, but it is also our responsibility as researchers to inform about what are and what are not the objectives of voice anonymization. We should ensure that our technology is unsuitable for generating deepfakes such that anonymized audios should always be recognizable as manipulated audios.

\section{Conclusion}
In this paper, we presented the to our knowledge first analysis of use cases for voice anonymization. Through an extensive literature and user study, we examined the differences between what users expect or want from an anonymization and what the current research focuses on. We find that the people from the general public have more interest in anonymization in scenarios that involve human-human interactions such as online communication and legal testimonies, while the research focuses more on situations involving human-computer interfaces. From the outcome of this use case collection, we derive a taxonomy of 25 use cases in voice anonymization, divided into applications of human-human and human-computer interaction, as well as data storage. We identify different requirements to the anonymization depending on the use case, and define criteria that need to be considered during system development and evaluation in order to meet these requirements. We hope that this scheme will help researchers in making decisions for developing new methods and to support open communication about assumptions and limitations of voice anonymization systems.

\section{Acknowledgements}
This work is funded by the Deutsche Forschungsgemeinschaft (DFG, German Research  Foundation) – Project: Multilingual Controllable Voice Privacy (VoiPy) - Project number 533241795.

\bibliographystyle{IEEEtran}
\bibliography{mybib}

\end{document}